\documentclass[11pt]{aastex}
\usepackage{graphicx,emulateapj5,apjfonts}
\usepackage{onecolfloat}
\usepackage{epsf}
\usepackage[hyperindex,breaklinks]{hyperref}

\shortauthors{Bell et al.}
\shorttitle{The stellar halo of the Milky Way}

\slugcomment{{\sc The Astrophysical Journal:} to appear June 20, 2008}

\begin{document}


\def\head{

\title{The accretion origin of the Milky Way's stellar halo}

\author{Eric F.\ Bell$^1$, Daniel B.\ Zucker$^2$, 
Vasily Belokurov$^2$, Sanjib Sharma$^3$, Kathryn V.\ Johnston$^3$, 
James S.\ Bullock$^4$, David W.\ Hogg$^5$, Knud Jahnke$^1$, 
Jelte T.\ A.\ de Jong$^1$, Timothy C. Beers$^{6}$, N.\ W.\ Evans$^2$, 
Eva K.\ Grebel$^{7,8}$, \v{Z}eljko Ivezi\'c$^9$, Sergey E.\ Koposov$^1$, 
Hans-Walter Rix$^1$,  
Donald P.\ Schneider$^{10}$, Matthias
Steinmetz$^{11}$, and Adi Zolotov$^5$
}
\affil{$^1$ Max-Planck-Institut f\"ur Astronomie,
K\"onigstuhl 17, D-69117 Heidelberg, Germany; \texttt{bell@mpia.de}\\
$^2$ Institute of Astronomy, University of Cambridge, Madingley
Road, Cambridge CB3 0HA, UK \\
$^3$ Department of Astronomy, Columbia University, 550 West 120th Street, 
New York, NY 10027, USA \\
$^4$ Center for Cosmology, Department of Physics and Astronomy, 
University of California, Irvine, CA 92697, USA \\
$^5$ Center for Cosmology and Particle Physics,
Department of Physics,
New York University,
4 Washington Place \#424,
New York, NY 10003, USA \\
$^6$ Department of Physics and Astronomy, Center for the Study of Cosmic Evolution and Joint
Institute for Nuclear Astrophysics, Michigan State University, E.\ Lansing, MI 48824, USA \\
$^7$ Astronomical Institute, Department of Physics \& Astronomy,
University of Basel, Venusstrasse 7, CH-4102 Binningen, Switzerland \\
$^8$ Astronomisches Rechen-Institut, Zentrum f\"ur Astronomie, 
Universit\"at Heidelberg, 
M\"onchhofstr.\ 12--14, D-69120 Heidelberg,
Germany \\
$^9$ University of Washington, Dept.\ of Astronomy, Box 351580, Seattle, WA 98195, USA \\
$^{10}$ Department of Astronomy and Astrophysics, 
The Pennsylvania State University, University Park, PA, USA \\
$^{11}$ Astrophysical Institute Potsdam, An der Sternwarte 16, D-14482 Potsdam, Germany }

\begin{abstract}
We have used data from the Sloan
Digital Sky Survey (SDSS) Data Release 5 to explore the overall 
structure and substructure
of the stellar halo of the Milky Way using $\sim 4$ million color-selected
main sequence turn-off stars with $0.2 < g-r < 0.4$ and $18.5 \le r < 22.5$.  
We fit oblate and triaxial broken
power-law models to the data, and found a `best-fit' 
oblateness of the stellar halo $0.5<c/a<0.8$, and 
halo stellar masses between Galactocentric radii of 1 and 40\,kpc of
$3.7 \pm 1.2 \times 10^{8} M_{\sun}$.  The density profile of 
the stellar halo is approximately $\rho \propto r^{-\alpha}$, 
where $-2 > \alpha > -4$.  Yet, we found 
that all smooth and symmetric models were 
very poor fits to the distribution of stellar halo stars because
the data exhibit a great deal of spatial substructure.  
We quantified deviations from a smooth oblate/triaxial model 
using the RMS of the data around the model 
profile on scales $\ga 100$\,pc, after 
accounting for the (known) contribution of Poisson uncertainties.
Within the DR5 area of the SDSS, the fractional RMS deviation
$\sigma$/total of the actual
stellar distribution from any smooth, parameterized halo model
is $\ga 40$\%: hence, the stellar halo is highly structured.
We compared the observations
with simulations of galactic stellar halos formed entirely
from the accretion of satellites
in a cosmological context by analyzing the simulations in the 
same way as the SDSS data.  
While the masses, overall profiles, and degree of substructure 
in the simulated stellar halos show considerable scatter,
the properties and degree of substructure in 
the Milky Way's halo match well the properties of 
a `typical' stellar halo built exclusively out of the debris
from disrupted satellite galaxies.  
Our results therefore point towards a picture in which an important
fraction of the stellar halo of the Milky Way has been accreted from
satellite galaxies.
\end{abstract}

\keywords{Galaxy: halo --- Galaxy: formation --- Galaxy: evolution ---
galaxies: halo --- Galaxy: structure --- Galaxy: general}
}

\twocolumn[\head]

\section{Introduction}

The stellar halo of the Milky Way has a number of 
distinctive characteristics which make it a key probe
of galaxy formation processes.  Milky Way halo stars
have low metallicity, alpha element enhancement, 
a high degree of support from random 
motions, and a roughly $r^{-3}$ power 
law distribution in an oblate
halo \citep{els,chiba00,Ya00,larsen03,lemon04,newberg,Ju06}.  
The low metallicities 
and alpha element enhancements 
suggest that the {\it stars} formed 
relatively early in the history of 
the Universe.  Yet, there has been 
disagreement about where these stars formed:
did they form {\it in situ} in the early phases
of the collapse of the Milky Way \citep[e.g.,][]{els}, or did the stars
form outside the Milky Way in satellite galaxies 
only to be accreted by the Milky Way at a later date 
\citep[e.g.,][]{sz,majewski96,bkw,bullock,abadi06}?

A key discriminant between these pictures 
is the structure of the stellar halo \citep{majewski93}.  {\it In situ}
formation would predict relatively little substructure, 
as the formation epoch was many dynamical times ago.
In contrast, current models of galaxy formation in a hierarchical
context predict that the vast majority of 
stellar halo stars should be accreted from disrupted satellite 
galaxies \citep{johnston98,bkw,bullock,moore06,abadi06}.  
The accumulated debris from ancient 
accretion episodes rapidly disperses in real 
space \citep[although in phase space, some information about
initial conditions persists; e.g.,][]{helmi99}, forming a relatively
smooth stellar halo.  The debris from accretions in 
the last few Gyr can remain in relatively distinct 
structures.  Simulations predict
quite a wide range in `lumpiness' of stellar halos, 
with a general expectation of a significant 
amount of recognizable halo substructure \citep{bkw,bullock}.  

Consequently, a number of groups have searched for
substructure in the Milky Way's stellar halo, 
identifying at least 3 large-scale 
features --- tidal tails from the disruption of 
the Sagittarius dwarf galaxy, the Low-Latitude stream, and 
the Virgo overdensity 
\citep[although see Momany et al.\ 2006 for a discussion of possible disrupted disk origin of much of the Low-Latitude stream]{ibata95,Ya00,iv00,newberg02,majewski03,yanny03,ibata03,Ju06,duffau06,belokurov06_sgr,newberg_new} --- and a host of 
tidal tails from globular clusters 
\citep[e.g.,][]{odenkirchen03,grillmair06_ngc}, dwarf
galaxies \citep[e.g.,][]{ih,md01}, and of unknown 
origin \citep[e.g.,][]{belokurov06_orp,grillmair06_orphan,grillmair06_cold,belokurov07_her}.  
Furthermore, substructure has been observed
in the stellar halos of other galaxies \citep[e.g.,][]{shang98,ibata01}.
Thus, it is clear that accretion of stars from satellite galaxies
is a contributor to the stellar halos of 
galaxies.  

Yet, it remains unclear whether
accretion is the {\it dominant} mechanism for
halo build-up.  A key observable is the 
fraction of stars in substructure (or a quantitative measure of the degree of 
substructure): if much of 
the halo mass is held in substructures, this argues 
for an accretion origin; if instead a tiny fraction of halo stars
is held in substructures, this places (very) tight constraints
on any recent accretion scenario.  However, it is not clear
how best to address this question.  One possible
approach is to define `overdense' areas of the halo by hand or 
algorithmic means, and to fit the rest with a smooth halo 
component; the remainder would be in `overdensities' 
\citep[e.g.,][]{newberg}.  
Here, we take a different approach.  Since one does 
not know {\it a priori} which stars should be `smooth halo'
stars and which are in `overdensities', we treat all halo
stars equally, fit a smooth model, and examine the 
RMS of the data around that smooth model (accounting
for the contribution to the RMS from counting statistics).
In this way, we obtain a quantitative measure
of the degree of halo structure on $\ga 100$\,pc scales without having to make
uncomfortable decisions about which stars should be fit with 
a smooth component and which should be included in overdensities.

In this paper, we apply this technique to 
explore the structure of the stellar
halo of the Milky Way, and place constraints on the 
fraction of stars in stellar halo under- or over-densities 
using imaging data from the 
Fifth Data Release (DR5) of the 
Sloan Digital Sky Survey \citep[SDSS;][]{Yo00,Am07}.  
Under the assumption that the bulk of the stellar population 
in the stellar halo is relatively metal-poor and old, we
isolate a sample dominated by halo main sequence turn-off stars
and explore the distribution of halo stars as a function of 
Galactic latitude, longitude and distance from the Sun (\S \ref{sec:data}).
In \S \ref{sec:models}, we generate a grid of smooth halo models
to compare with the data, and in \S \ref{sec:results} we 
constrain the `best-fit' smooth stellar halo parameters and 
quantify the fraction of halo stars in stellar halo 
under- or over-densities.  We compare the observations
with models of stellar halo formation in a cosmological context in 
\S \ref{sec:disc}, and present a summary 
in \S \ref{sec:conc}.

\section{Data}
\label{sec:data}

SDSS is an imaging and spectroscopic survey that has
mapped $\sim 1/4$ of the sky. Imaging data are produced simultaneously
in five photometric bands, namely $u$, $g$, $r$, $i$, and
$z$~\citep{Fu96,Gu98,Ho01,Gu06}. The data are processed through
pipelines to measure photometric and astrometric properties
\citep{Lu99,St02,Sm02,Pi03,Iv04,tucker06} and to select targets for
spectroscopic follow-up.  DR5 covers $\sim 8000$ square degrees around the
Galactic North Pole, together with 3 strips in the Galactic southern
hemisphere. We use the catalog of objects classified as stars with
artifacts removed~\footnote{See
\url{http://cas.sdss.org/astro/en/help/docs/realquery.asp\#flags}}, together
with magnitude limits $ r < 23.5$ and $g < 24.5$.  Photometric 
uncertainties as a function of magnitude are discussed in
\citet{sesar07}.  We choose to 
analyze only the largely contiguous $\sim 8000$ square degree area
around the Galactic North Pole in this work, giving a total sample of 
$\sim 5 \times 10^7$ stars, of which $\sim 3.6 \times 10^6$ stars
meet the selection criteria we apply later.  
In what follows, we use Galactic extinction 
corrected magnitudes and colors, following \citet{Sc98}; 
such a correction is appropriate for the stars of interest in this paper
owing to their large heliocentric distances $D_{\rm heliocentric} \ga 8$\,kpc.

\subsection{Color--magnitude diagrams: an introduction}
\label{sec:cmds_intro}

\begin{figure*}[t]
\begin{center}
\epsfxsize 18.0cm
\epsfbox{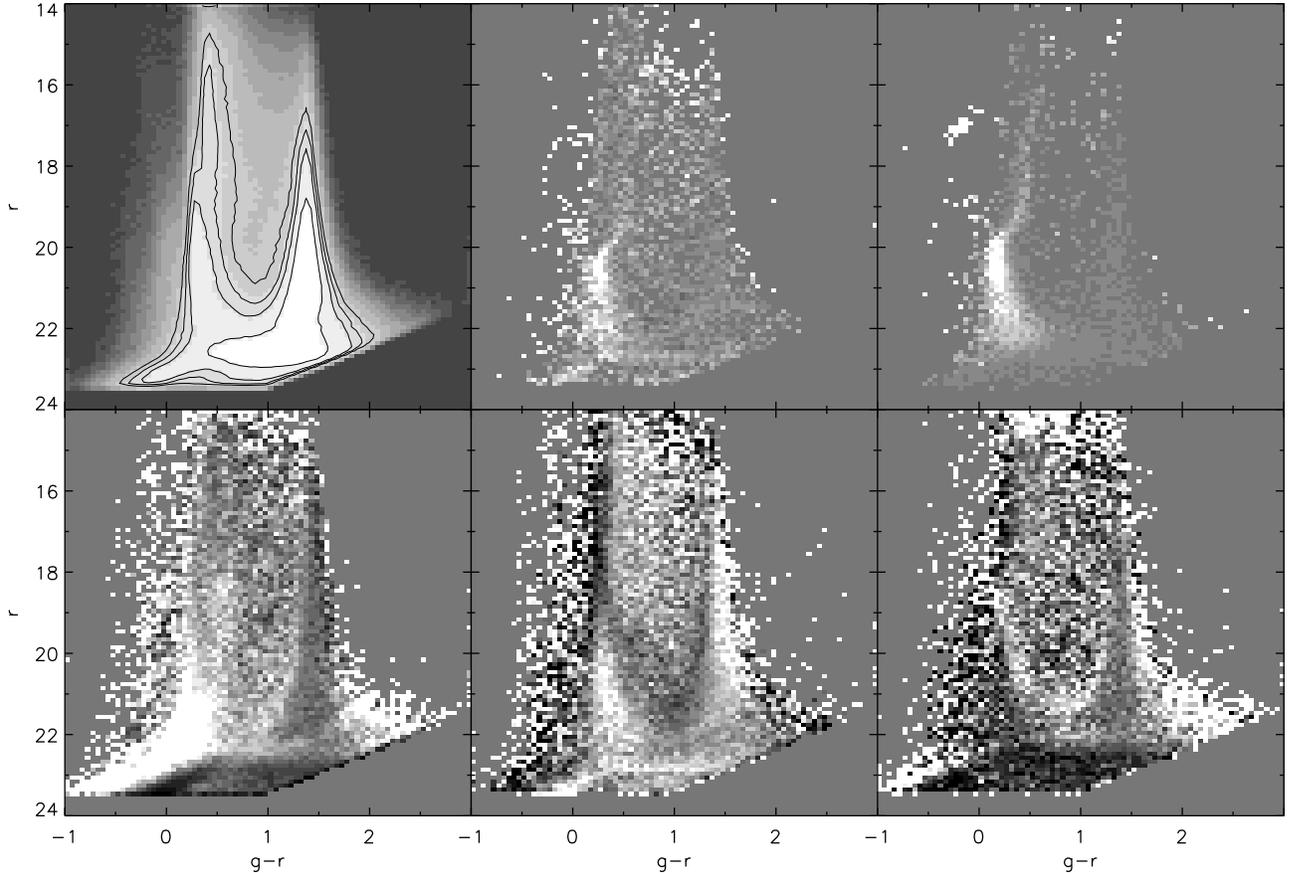}
\end{center}
\caption{\label{fig:cmd}  
Hess diagrams in terms of $g-r$ color and $r$-band magnitude derived from 
the SDSS data.  In these Hess diagrams, we show for completeness the 
data to the very faintest limits $r \ga 23$, where the 
S/N is low and there is significant contamination 
by misclassified galaxies, spurious detections, etc.  These
diagrams show in general two plumes in the stellar density
distribution that reflect main sequence turn-off stars with 
$g-r \sim 0.3$ and intrinsically faint and low-mass disk 
dwarf stars with $g-r \sim 1.4$.  We limit
our analysis to $18.5 \le r < 22.5$ in this paper for the 
main sequence turn-off dominated color bin $0.2< g-r < 0.4$, 
in the area where the data quality
is still excellent.  
{\bf Top left:} The density of stars per square degree per color
interval per magnitude 
for $b>30\arcdeg$, scaled logarithmically.  This Hess diagram contains
$4 \times 10^7$ stars. 
{\bf Top middle:} The Hess diagram
for the (sparsely-populated) globular cluster Pal 5 (within a circle
of radius $0\fdg5$).
{\bf Top right:} The Hess diagram
for the globular cluster NGC 5024.  {\bf Bottom left:} 
A difference Hess diagram (following Eqn.\ \ref{eqn:diff})
differencing two lines of sight $(l,b) = (300,70)$ and
$(l,b) = (60,70)$.  The grey scales saturate at $\pm 100$\%.
In an axisymmetric halo, this difference should
equal zero within the shot noise: it clearly does not.  {\bf Bottom middle:} 
A difference Hess diagram 
differencing two lines of sight $(l,b) = (44,40)$ and
$(l,b) = (15,45)$.  The grey scales saturate at $\pm 50$\%.
This Hess diagram should be close to, but not exactly equal
to, zero.
{\bf Bottom right:} 
A difference Hess diagram 
differencing two lines of sight $(l,b) = (167,35)$ and
$(l,b) = (193,35)$.  The grey scales saturate at $\pm 50$\%.
Again, in a symmetric halo, this difference should
equal zero.}
\end{figure*}

To help get one's bearings, it is instructive to examine some 
color--magnitude diagrams (CMD) derived from these data (Fig.\ \ref{fig:cmd}).
The color--magnitude diagram of all stars with $b > 30\arcdeg$ is 
shown in the top left panel, where the grey levels show the 
logarithm of the number of stars in that bin per square degree
from $10^{-3}$ stars/deg$^2$ to $7.1$ stars/deg$^2$; such a 
scaled CMD is frequently called a Hess diagram.
To help interpret this Hess diagram, we show two 
additional Hess diagrams for two globular clusters covered
by these data: Pal 5 and NGC 5024 \citep[in what follows, 
distances and metallicities for these and all other globular clusters
are adopted from][]{harris96}.
The top middle panel of Fig.\ \ref{fig:cmd} shows a Hess diagram for
stars in the globular cluster Pal 5 (a circle of radius 
$0\fdg5$ around the position $l=0\fdg85$ 
and $b=45\fdg9$).  The grey levels show: 
\begin{equation}
(N_{\rm on}A_{\rm on}^{-1} - N_{\rm off}A_{\rm off}^{-1})/N_{\rm off}A_{\rm off}^{-1}, \label{eqn:diff}
\end{equation}
where $N$ denotes the number of stars in the field of interest
(denoted by the subscript `on') and a control field `off', and $A$ is 
the area of that field.  In this case the control field is nearby:
a circle of radius $4\arcdeg$ around the position $l=6\arcdeg$ 
and $b=46\arcdeg$.  One can clearly see the main sequence turn off
with $g-r \sim 0.2$ and $r \sim 21$, with the lower main sequence
extending redwards towards fainter magnitudes and the subgiant branch 
extending redwards towards brighter magnitudes.  In the top right
panel we show a similar Hess diagram for NGC 5024; because
this cluster is rather brighter than Palomar 5 the CMD is better populated
and shows a more prominent red giant branch (extending towards 
brighter magnitudes with $g-r \sim 0.5$) and horizonal 
branch (with $g-r \la 0$ and $r \sim 17$).

\begin{figure}[t]
\begin{center}
\epsfxsize 9.0cm
\epsfbox{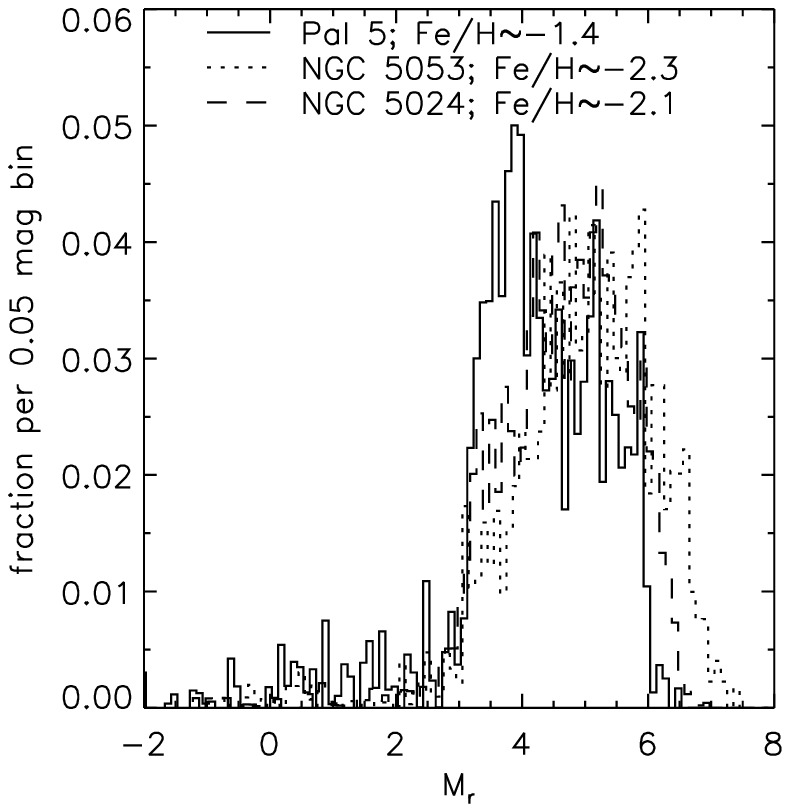}
\end{center}
\caption{\label{fig:hist}
A histogram of the absolute magnitudes 
of stars with $0.2<g-r<0.4$ in three
globular clusters: Pal 5, NGC 5053, and NGC 5024.  These
distributions give an impression of the actual convolution kernel
suffered by the $0.2<g-r<0.4$ 
MSTO stars in the halo of the Milky Way when going from distance
to apparent magnitude.  In this work, we choose to approximate
this distribution for modeling the stellar halo with a 
Gaussian distribution with $\langle M_r \rangle = 4.5$\,mag and 
$\sigma_{M_r} = 0.9$\,mag, an appropriate choice for a stellar 
population with [Fe/H]$\sim -1.5$.  
}
\end{figure}  

There are a few points to note about Fig.\ \ref{fig:cmd}.
Firstly, for old populations such as those in globular clusters
it is clear that the color
of the main sequence turn-off (MSTO) is a metallicity 
indicator (this point is discussed
in more detail for SDSS isochrones in \citealp{girardi04}).  Comparing Pal
5 (${\rm [Fe/H] \sim -1.4}$, $(g-r)_{\rm MSTO} \sim 0.3$) 
with NGC 5024 (${\rm [Fe/H] \sim -2.1}$, $(g-r)_{\rm MSTO} \sim 0.15$), 
one can see that old very metal-poor populations (${\rm [Fe/H] \la -2}$)
have bluer main 
sequence turn-offs compared to less metal-poor populations  
(${\rm [Fe/H] \sim -1.5}$).  Second, MSTO stars are a reasonably
good distance indicator, albeit with significant scatter.  In 
Fig.\ \ref{fig:hist}, we show the absolute magnitude distribution of 
all stars with $0.2<g-r<0.4$ in Pal 5 (solid line: distance$=22.6$\,kpc), 
NGC 5024 (dashed line: distance$=18.4$\,kpc) 
and a third globular cluster
NGC 5053 (dotted line: ${\rm [Fe/H] \sim -2.3}$, distance$=16.2$\,kpc).  
The mean $r$-band absolute magnitudes 
of the distributions are (4.3,4.7,5.0) respectively, 
and all distributions individually have RMS $\sim 0.9$ mag.
Thus, modulo a metallicity-dependent $\la 0.5$ systematic
uncertainty, the MSTO is a good distance indicator
with $\sim 0.9$ mag scatter.

Examining the top left panel of Fig.\ \ref{fig:cmd}, in the 
light of the globular cluster CMDs, it is possible to interpret
some of the features of the $b>30\arcdeg$ Hess diagram.
At all distances, the MSTO is visible as a clearly-defined
as a sharp `blue edge' to the distribution, indicating to 
first order that the stars in the galactic disk at large scale 
heights and in the stellar halo are dominated by a metal-poor old population 
with ages not that dissimilar to those of globular clusters; 
this is the assumption that we will adopt in the remainder of this paper.
At $g-r < 0.5$, one sees the MSTO for stars in the stellar disk at 
$\ga$kpc scale heights
(at $r<18$; often the disk at such scale heights is referred to 
as the thick disk) and in the stellar halo (at $r>18$).  One can 
see a `kink' in the MSTO at $r \sim 18$, as highlighted by the contours; 
we interpret
this as signifying a metallicity difference between the 
disk at $\sim$\,kpc scale heights and stellar halo (this transition 
is also very prominent in Fig.\ 4 of \citealp{lemon04} and in 
\citealp{chen01}, who interpret
this CMD feature in the same way).  The MSTO in the stellar halo
has $g-r \sim 0.3$, similar to that of Pal 5 (${\rm [Fe/H] \sim -1.4}$)
and $\sim 0.15$\,mag redder than those of NGC 5024 and NGC 5053 
with (${\rm [Fe/H] \la -2}$).  This suggests a halo 
metallicity ${\rm [Fe/H] \sim -1.5}$, in excellent 
agreement with measured halo metallicity distributions, which 
peak at  ${\rm [Fe/H] \sim -1.6}$
\citep[e.g.,][]{laird98,venn04}.

\subsection{Hess diagrams of structure in the stellar halo}
\label{sec:cmds_struc}

One of the main goals of this paper is to explore the 
degree of substructure in the stellar halo of the Milky Way.
One way of visualizing this issue is through the inspection of 
Hess diagrams where pairs of lines of sight 
are subtracted, following Eqn.\ \ref{eqn:diff}\footnote{An extension 
of this methodology was used by \citet{xu06}, who use the SDSS
DR4 to study stellar halo structure using star counts
and color distributions of stars at Galactic latitudes $b \ge 55\arcdeg$.}.  
We have done this exercise for 
three such lines of sight in Fig.\ \ref{fig:cmd}, 
where we have chosen three line-of-sight pairs where
the subtraction {\it should} have been close
to zero, if the stellar halo of the Milky Way were symmetric and smooth.

The lower left panel of Fig.\ \ref{fig:cmd} 
shows the difference of two different
lines of sight $(l,b) = (300,70)$ --- a line of sight
towards the Virgo overdensity and a part of the 
Sagittarius stream --- and  $(l,b) = (60,70)$; 
in a symmetric model such a subtraction should come out
to zero.  The grey levels saturate at deviations of 
$\pm 100$\%.  It is clear that the 
$(l,b) = (300,70)$ line-of sight has strong order-of-unity 
overdensities at 
MSTOs fainter than $r>21$, or distances of $>20$\,kpc
assuming a MSTO absolute magnitude of $M_r \sim 4.5$. 
One can see also a weak sub-giant and red giant branch feature
at $g-r \sim 0.5$ and $18<r<20$, again indicating 
distances $>20$\,kpc.

The lower middle shows a line of sight towards
$(l,b) = (44,37)$ 
minus the Hess diagram for stars towards $(l,b) = (15,41)$.
This subtraction would be expected to come out close
to, but not exactly, zero.  
It would be ideal to be able to subtract off 
the `correct' pairing of $(l,b) = (316,37)$; however, 
SDSS has not mapped that area of sky owing to 
its southern declination, $\delta = -25$.  The grey scale saturates
at $\pm 50$\%.  There are
minor artifacts in the subtraction; however, one can 
clearly see an overdensity of main sequence stars with 
a MSTO with $r \sim 20.5$, corresponding to a distance of 
$\sim 16$\,kpc.  

The lower-right panel shows a line of sight towards 
$(l,b) = (167,35)$ --- a line of sight towards part of the Low-Latitude
overdensity --- minus that of $(l,b) = (193,35)$.
In a symmetric halo this subtraction should be identically zero.  
The grey scale saturates at $\pm 50$\%.  There is a weak 
MSTO overdensity at $r \sim 19$ mag, some $\sim 7$\,kpc from 
the observer.  

While these lines of sight have been selected to show (varying 
degrees) of halo inhomogeneity\footnote{Although, in fact, we found it
impossible to avoid at least low-level inhomogeneity along 
any pair of lines of sight.}, they suffice to illustrate two
key points.  First, the halo is far from homogeneous, with strong
order-of-unity overdensities as well as weaker $\sim 10-20$\% features.
Second, owing to the partial sky coverage of the SDSS, it is 
difficult to visualize and quantitatively explore the structure of the 
Milky Way's stellar halo using CMD subtractions.  

\begin{figure*}[thb]
\begin{center}
\hspace{-1.cm}
\epsfxsize 19.0cm
\epsfbox{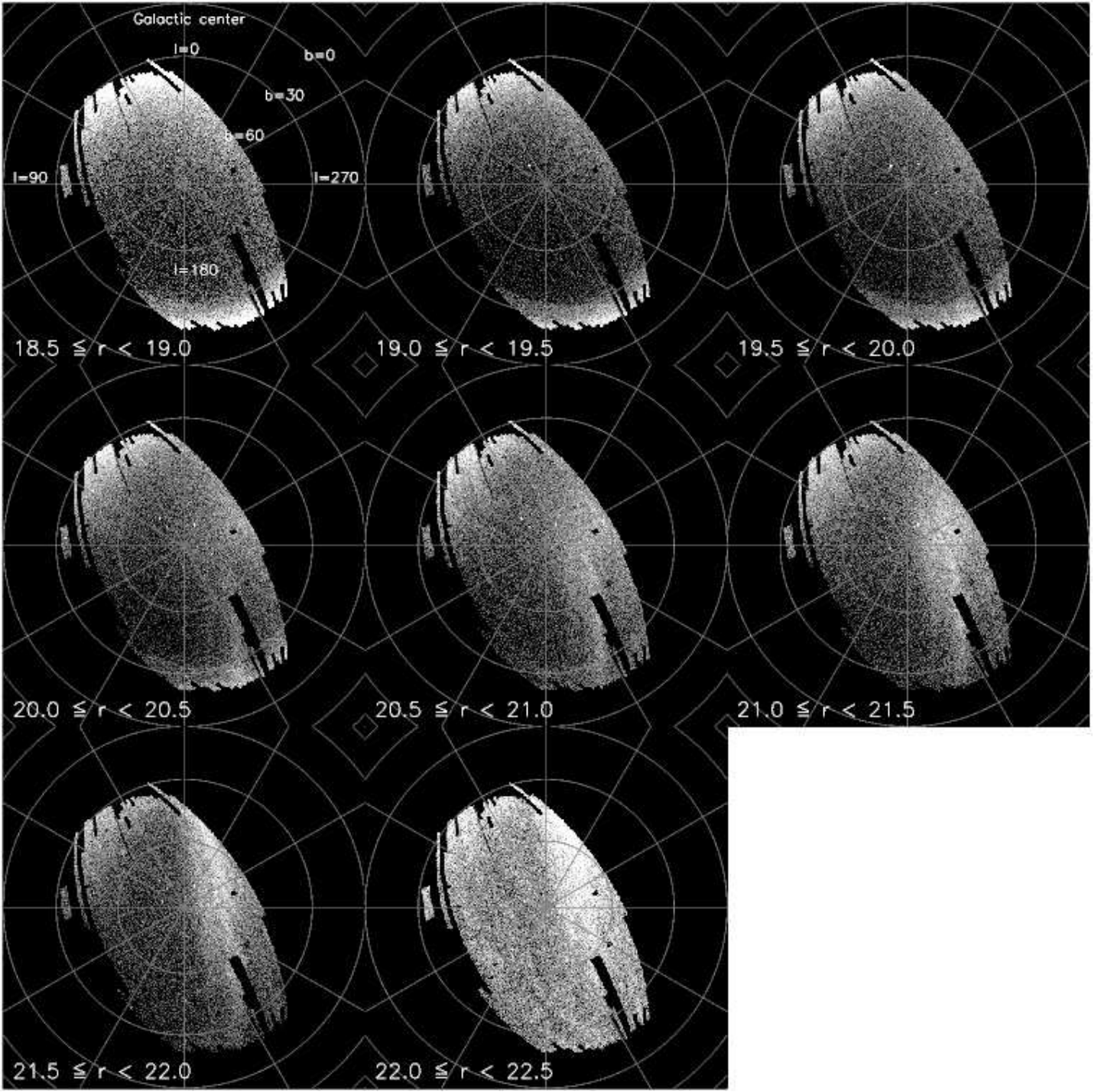}
\end{center}
\caption{\label{fig:halo} 
The stellar halo of the Milky Way as seen by SDSS.  The grey 
scale denotes the logarithm of the number density of 
$0.2 \le g-r \le 0.4$ stars 
per square degree in eight different magnitude (therefore mean distance)
slices; a Lambert azimuthal equal-area polar projection is used. 
The black areas are not covered by the SDSS DR5, and reflect
the great circle scanning adopted by the SDSS when collecting its 
imaging data.  Apparent `hot pixels' are stellar overdensities from 
globular clusters and dwarf galaxies.
}
\end{figure*}

\subsection{Main sequence turn-off star maps of the stellar halo}
\label{sec:maps}

One more intuitive approach to the distribution of stars
in the stellar halo is to construct maps of the number of MSTO
stars in different magnitude (therefore, roughly distance) slices.
We select MSTO stars with foreground extinction-corrected
$0.2<g-r<0.4$; this color range was selected empirically to 
encompass the most densely-populated bins of color space for 
the halo MSTO stars in Fig.\ \ref{fig:cmd}.  In \S \ref{sec:cmds_intro}, 
we showed that in such a color bin the average absolute magnitude
of the MSTO stars in that bin were 4.3 and 4.7 respectively for 
Palomar 5 (${\rm [Fe/H] \sim -1.4}$) and NGC 5024 (${\rm [Fe/H] \sim -2.1}$);
accordingly, we adopt an average MSTO $M_r =4.5$ in what follows for
stars in the color bin $0.2<g-r<0.4$.  Such an absolute magnitude
is in agreement with model CMDs, which suggest $M_r = 4.7\pm0.2$ for 
stars with metallicities ${\rm [Fe/H] \sim -1.5\pm0.5}$.  We make the assumption
that all stars in the stellar halo are `old' (i.e., approximately the same
age as the calibrating globular clusters).

We show 0.5\,mag wide bins of $r$-band magnitude
between $18.5 \le r < 22.5$, corresponding to heliocentric distances
of $7 \la d/{\rm kpc} \la 40$.    At such heliocentric distances, the 
vertical distance above the Galactic plane
is $\ga 5$\,kpc along all lines of sight, or at $\ga 5$ scale 
heights following the $\sim 900$\,pc thick disk scale height estimated
by \citet{larsen03}.  Thus, the dominant contribution to the 
MSTO maps is from the stellar halo.
The resulting Lambert azimuthal 
equal-area polar projections, logarithmically-scaled, 
are shown in Fig.\ \ref{fig:halo}\footnote{This presentation 
is similar to that of Fig.\ 24 of \citet{Ju06}, who present
this kind of analysis for $20 < r < 21$, and \citet{newberg_new}, 
who present a similar diagram with slightly more restrictive 
color cuts for $20 < g < 21$.}.

While one loses the ability to probe for population differences in the stellar
halo because of the broad color bin adopted to derive these
densities, it is 
much more straightforward to visualize the distribution of halo
MSTO stars using this technique.  From Figs.\ \ref{fig:cmd} and 
\ref{fig:hist}, one can see that MSTO stars at a single distance will 
show up in multiple distance bins: the bins are 0.5\,mag wide, and 
the RMS of a single distance stellar population is $\sim 0.9$\,mag.
This can be seen easily from inspection of some of the `hot pixels'
in Fig.\ \ref{fig:halo}, 
corresponding to known globular clusters or dwarf galaxies.
These features persist from map-to-map despite there being a single
population at a unique distance, giving a visual impression of the 
covariance between the different maps.

\clearpage

Focusing on the brightest bins, $18.5\le r <20$, corresponding to 
heliocentric distances between $\sim 7$\,kpc and $\sim 11$\,kpc, 
the stellar distribution appears rather smooth, with 
higher density towards the Galactic center and Galactic 
anticenter.  In the case of the Galactic center, the interpretation 
is straightforward: one is probing lines of sight which 
pass $\sim 5$\,kpc from the Galactic center, and probe the denser
inner parts of the stellar halo.  In the case of the Galactic anticenter, 
such a structure is not expected in a oblate/triaxial halo model, and 
recalling the $\la 1$\,kpc scale height of the thick disk 
cannot be a thick disk; 
this is the well-known Low-Latitude stream 
\citep[e.g.,][]{newberg02,pena05,momany06}.  
In this visualization, the stream appears to be spread out 
between a few different magnitude bins: at $b < 30\arcdeg$ some 
of that spread may be real, but the well-defined structure at 
$(l,b) \sim (165,35)$ has a relatively narrow distance spread (see
the Hess diagram residual 
in the lower right-hand panel of Fig.\ \ref{fig:cmd}, showing
a reasonably narrow main sequence;  see also the discussion 
in \citealp{grillmair06_mon}).  

Focusing on the more distant bins, $20 \le r < 22.5$, corresponding to 
heliocentric distances between $\sim 14$\,kpc and $\sim 35$\,kpc, one finds
little contribution from the Low-Latitude stream.  Instead, superimposed
on a reasonably smooth background is a prominent contribution from 
large tidal tails from the ongoing interaction of the Milky Way with the 
Sagittarius dwarf galaxy \citep[see][for a much more detailed discussion]{belokurov06_sgr}.  As quantified by \citet{belokurov06_sgr}, 
one can discern a distance gradient in 
the stream, from the closest populations
towards the Galactic anticenter $(l,b) \sim (200,20)$ to
the most distant populations towards $(l,b) \sim (340,50)$.  
 
While it is clear from these maps that the stellar halo of the 
Milky Way is not completely smooth, there is a `smooth' (i.e., not 
obviously structured) component
which dominates these maps: if there are variations
in this component, these must be on spatial scales 
$\ga 10\arcdeg$ on the sky (or scales $\ga 1$\,kpc at
the distances of interest for this paper).  A number of methods
could be devised to probe halo structures on such scales.  In this paper, 
we choose to construct models of a smooth stellar halo to represent
the Milky Way, and to ask about the fraction of stars
deviating from this smooth global model as a measure of substructure
in the halo.  This exercise is the topic of the remainder of this paper.

\section{Models of a Smoothly-Distributed Stellar Halo}
\label{sec:models}

The stellar halo of the Milky Way is modeled using an  
triaxial broken power-law, where we explore oblate and prolate 
distributions as special cases of triaxial.  The minor axis of the 
ellipsoid is constrained to be aligned with the normal to the 
Galactic disk (this is is contrast with \citealp{newberg} and 
\citealp{xu06}, who allow 
the minor axis to vary freely).
There are 7 free parameters: the normalization 
$A$ (constrained directly through requiring that the model and 
observations have the same number of stars in the magnitude and 
color range considered in this paper), the two power
laws $\alpha_{\rm in}$ and $\alpha_{\rm out}$, the break
radius $r_{\rm break}$, $b/a$, $c/a$, and the Galactocentric
longitude of the major axis $L_{\rm major}$.  We adopt a grid search, 
with between 4 and 10 values in each parameter of interest, yielding 
typically several hundred to several thousand smooth models to test 
against the data.  
In what follows,
we assume a distance to the Galactic center of 8\,kpc and 
a $M_r = 4.5$ for the MSTO stars with $0.2<g-r<0.4$, with 
a $\sigma_{M_r} = 0.9$\,mag.  
  
We choose to define the best fit to be the fit for which 
the RMS deviation of the data $\sigma$ around the model is minimized, 
taking account of the expected Poisson counting uncertainty in the model, 
summed over all bins in $l$, $b$, and magnitude:
\begin{eqnarray}
\langle \sigma^2 \rangle & = & \frac{1}{n} \sum_i (D_i-M_i)^2 - \frac{1}{n} \sum_i (M_i'-M_i)^2 \\
\sigma / {\rm total} &= & \frac{\sqrt{\langle \sigma^2 \rangle}}{\frac{1}{n} \sum_i D_i},
\end{eqnarray}
where $D_i$ is the observed number of main-sequence turn-off stars
in bin $i$, 
$M_i$ is the exact model expectation of that bin, 
$M_i'$ is a realization of that model drawn from a 
Poisson distribution with mean $M_i$, and $n$ is the number 
of pixels.   We could have chosen instead to define the best
fit by minimizing the reduced $\chi^2$:
\begin{equation}
\chi^2 = \sum_i (D_i - M_i)^2/\sigma_i^2, 
\end{equation}
where $\sigma_i^2$ is the Poisson uncertainty of the model $M_i$\footnote{The uncertainty in the model is chosen here because we are evaluating the likelihood of the dataset being drawn from the model.}.
We have chosen not to do so in this case because
we are interested in quantifying and placing a lower limit on the 
structure in the stellar halo in this paper, not in finding 
the `best' fit to the stellar halo in a $\chi^2$ sense (we show
in \S \ref{sec:results} that the  
stellar halo model with the lowest $\chi^2$ 
has a $\sigma$/total that is close to, but slightly 
higher than the stellar halo model with lowest $\sigma$/total).  For the 
purposes of substructure quantification, $\sigma$/total has two decisive 
advantages.
Firstly, unlike $\chi^2$, $\sigma$/total is {\it independent} of pixel
scale,\footnote{The quantity $\sqrt{\langle \sigma^2 \rangle}$ is inversely proportional to $n$ in the presence of intrinsic structure in the dataset, as is the quantity $\frac{1}{n} \sum_i D_i$, thus making $\sigma$/total pixel scale independent.} provided that the substructure 
in the halo is well-sampled
by the chosen binning scale. 
Secondly, for $\sigma$/total, the contribution of 
Poisson noise to $\sigma$ has been removed, leaving only the 
contribution of actual halo structure to the variance\footnote{ 
We have confirmed by rebinning the data by factors of 16 in area that 
$\sigma$/total is indeed independent of pixel scale; thus, the dominant
contribution to the intrinsic structure of the stellar halo must
be on linear scales $\ga 400$\,pc.}.  Thus, even though we have adopted
a pixel size of $0.5\arcdeg \times 0.5\arcdeg$ in what follows 
(corresponding to $> 100$\,pc 
scales at the distances of interest), our results are 
to first order independent of binning scale (because empirically 
we find that the vast majority of the 
variance is contained on $\sim$\,kpc scales and greater).   
We defer to a future paper the 
exercise of understanding and interpreting the scale dependence of 
stellar halo substructure.  The main uncertainty
in the estimated values of $\sigma$/total is from the 
major contributions of a few large structures on the 
sky to $\sigma$/total, both through influencing the
`best fit' and through their direct contribution to the residuals.
Later, we attempt to quantify this uncertainty through 
exclusion of the most obvious substructures from consideration 
before fitting and estimation of $\sigma$/total.

The model parameters (including the normalization) 
give an estimate of the total number of stars in the halo.
We calculate the total number of stars contained in the 
model with Galactocentric radius $1 \le r_{GC}/{\rm kpc} \le 40$.
In order to interpret this value as a mass, it is necessary
to convert the number of $0.2<g-r<0.4$ stars into a mass
by calculating a mass-to-number ratio.  We adopt an 
empirical approach, following \citet{newberg}.  Given that the 
Pal 5 MSTO color seems to be a good match to the stellar halo
MSTO color, we use the mass of Pal 5 
\citep[$\sim 5000$\,M$_{\sun}$;][]{odenkirchen02} and 
the number of stars in Pal 5 above background ($\sim$1069 stars
with $0.2<g-r<0.4$) to define a mass-to-number ratio 
$\sim 4.7$M$_{\sun}$/MSTO star\footnote{\cite{koch04} find 
a deficit of low-mass stars in the central parts of 
Pal 5, suggesting that this ratio may be a lower limit.}.  
This ratio is in excellent agreement
with values derived using stellar population models
for populations with $[{\rm Fe/H}] \le -1.5$; these models
have values of $\sim 4$M$_{\sun}$/MSTO star.  

\begin{figure*}[t]
\begin{center}
\hspace{-1.cm}
\epsfxsize 19.0cm
\epsfbox{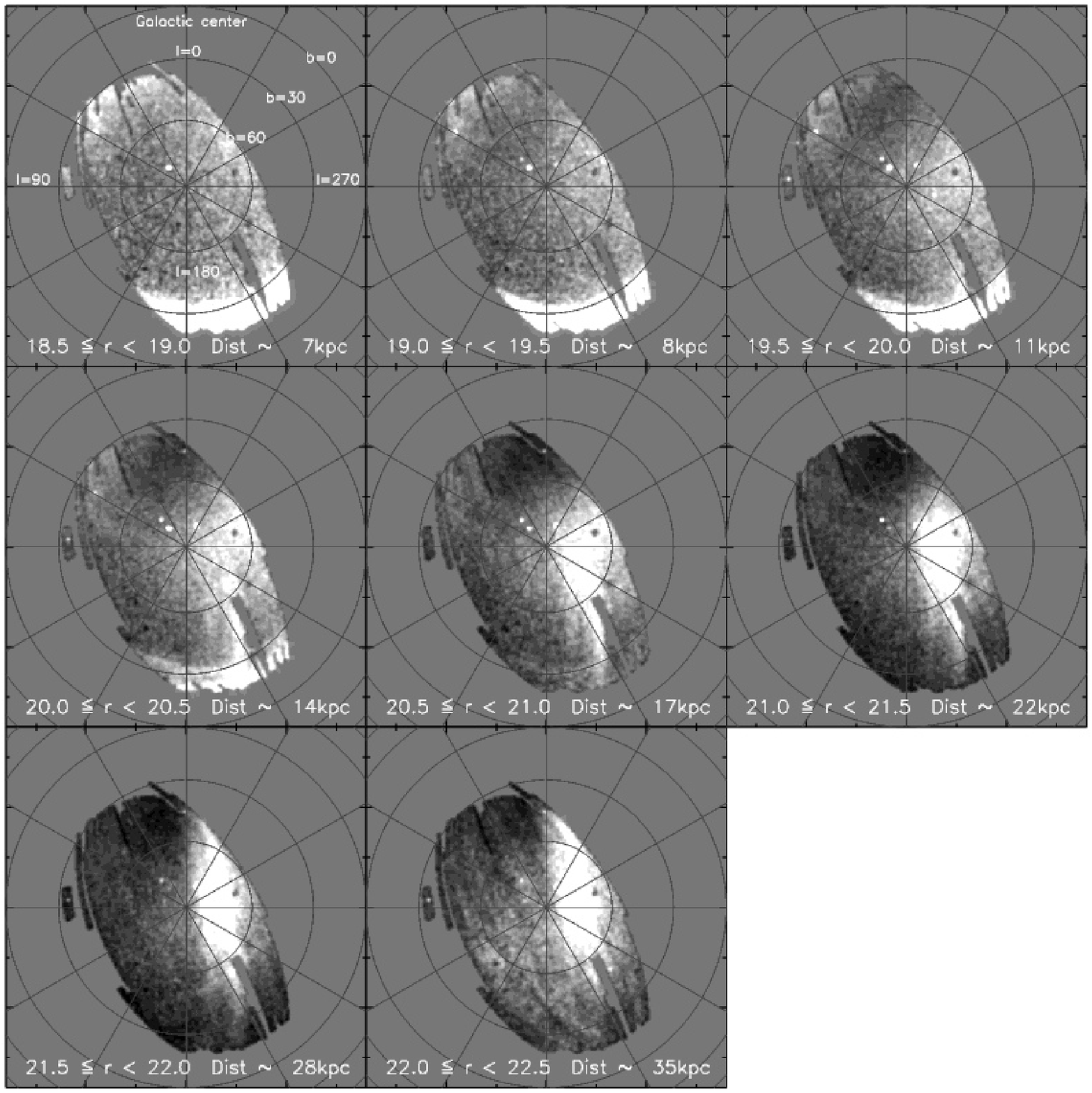}
\end{center}
\caption{\label{fig:resid} 
Residuals of the mean stellar density 
(data$-$model) from the best {\it oblate} model 
$(\alpha_{\rm in},\alpha_{\rm out},r_{\rm break},c/a) = 
(-2.2,-3.5,20\,{\rm kpc},0.7)$.  The panels show 8 different distance 
slices, and have been smoothed using a $\sigma = 42'$ Gaussian.   
The gray scale saturates at $\pm 60$\% deviation from the model density, 
and white represents an observed excess over the smooth model prediction.
}
\end{figure*}

As is clear from Figs.\ \ref{fig:halo} and \ref{fig:resid}, 
a significant part of the deviations from 
a smooth stellar halo is driven by the 
Sagittarius and Low-Latitude streams, and by the Virgo 
overdensity.  We therefore
run the whole minimization twice, once allowing all $b>30\arcdeg$ 
data to define the fit, and a second time masking out most of the
Sagittarius and Low-Latitude streams, and the Virgo overdensity, 
by masking regions with $b < 35\arcdeg$ and $0 < X < 30$,
where $X$ is the abscissa of the equal-area projection: 
$X = 63.63961 \sqrt{2(1-\sin b)}$.  This masking is done
to constrain the importance of these
larger structures in driving the model parameters 
and residual fraction.  

\section{Results}
\label{sec:results}

In this section, we present the fitting results for a large
set of smooth, symmetric stellar halos.  In \S \ref{sec:ob}
we present the results from oblate stellar halos (i.e., the two longest
axes have equal lengths).  In \S \ref{sec:tri}, we
discuss the fitting results for triaxial stellar halos (where
all three axes can have different lengths), comparing this 
general case to the case of an oblate halo.

\begin{figure}[t]
\begin{center}
\hspace{-0.5cm}
\epsfxsize 9cm
\epsfbox{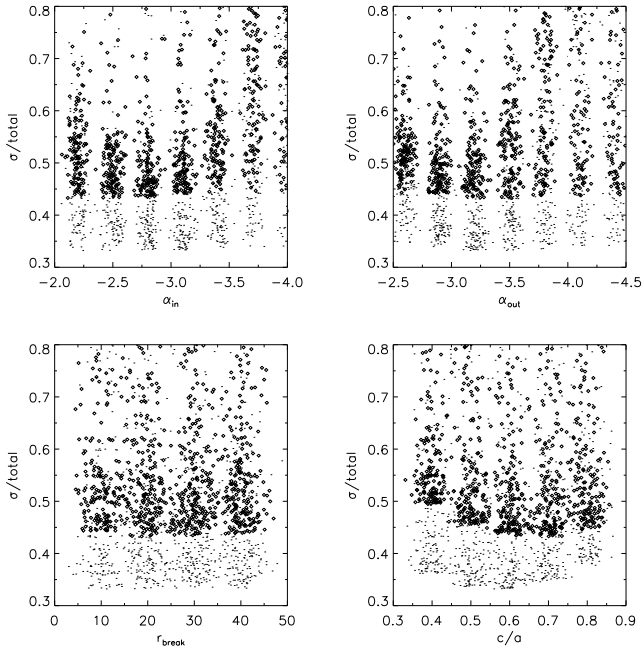}
\end{center}
\caption{\label{fig:para} 
The $\sigma$/total of a large number of {\it oblate} halo
models.  Each point represents the 
value of $\sigma$/total for a different oblate halo model:
open diamonds show the residuals when no clipping
is applied, points show the result when areas 
with contributions Sgr/Low-Latitude stream/Virgo
are excised before carrying out the analysis.  
In each case, we show the values of $\sigma$/total
as a function of $\alpha_{\rm in}$, $\alpha_{\rm out}$, 
$r_{\rm break}$ and $c/a$, marginalized over all other model
parameters.  Recall that our definition of $\sigma$/total 
subtracted off the Poisson uncertainties already, and 
is a measure of the degree of substructure on scales
$\ga 100$\,pc.  It is clear that the oblateness $c/a$ of the halo 
is the best-constrained parameter; combinations of all of the other parameters
can provide equally-good fits, given an oblateness.
Small random offsets are applied to the discrete values of 
$\alpha_{\rm in}$, $\alpha_{\rm out}$, $r_{\rm break}$ and $c/a$ to 
aid visibility.
}
\end{figure}

\begin{figure}[t]
\begin{center}
\hspace{-0.5cm}
\epsfxsize 9cm
\epsfbox{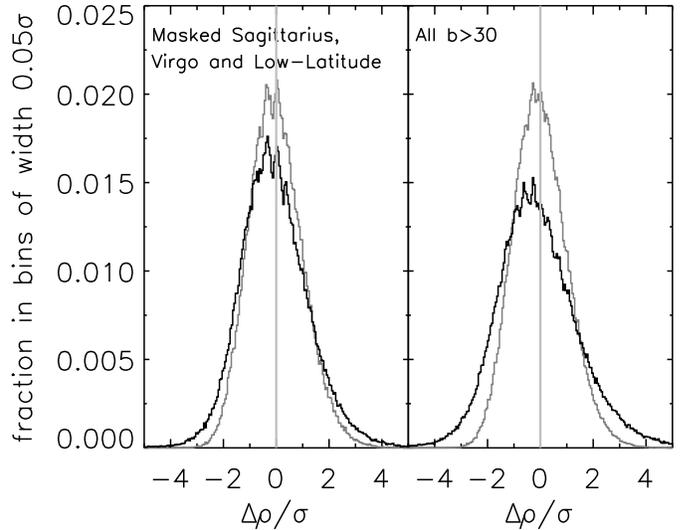}
\end{center}
\caption{\label{fig:hists} 
The distribution of differences between the observed star counts per
$0.5 \arcdeg \times 0.5 \arcdeg$ pixel and that predicted by the 
best-fitting smooth model, divided by the $\sigma$ predicted by 
Poisson uncertainties (black lines).  The gray line shows the expected
distribution from Poisson fluctuations around the smooth model.  The left
panel shows the distributions for the case in which sky areas of 
the Sagittarius, 
Virgo and Low-Latitude overdensities have been excised before this 
analysis; the right panel
shows the results for all $b>30\arcdeg$ data.  Note that $\sim 1/2$ of 
the excess variance is in the `peak' of the histogram 
(with $|\Delta \rho| < 3 \sigma$), 
and the rest of the excess variance reflects a number 
of pixels with $|\Delta \rho| > 3\sigma$ 
(predominantly towards overdensities, rather 
than towards underdensities).  
}
\end{figure}

\begin{figure}[t]
\begin{center}
\hspace{-0.5cm}
\epsfxsize 9cm
\epsfbox{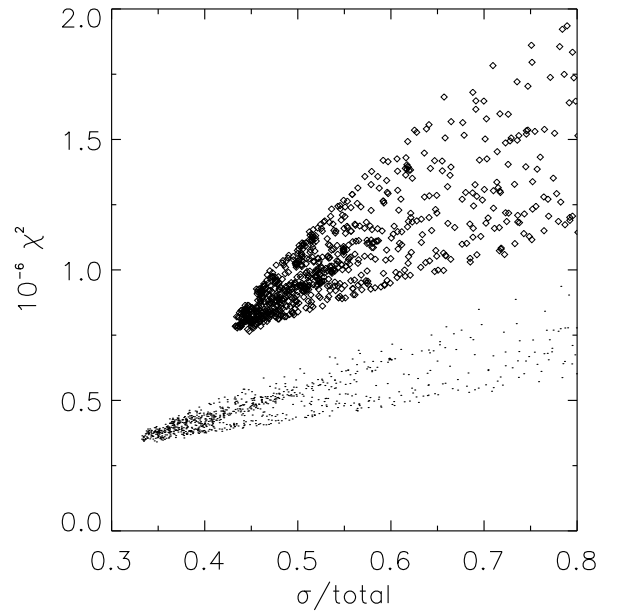}
\end{center}
\caption{\label{fig:chisig} 
A comparison between $\sigma$/total and $\chi^2$ for 
a large number of {\it oblate} halo
models.  Each point represents a different oblate halo model:
open diamonds show the residuals when no clipping
is applied (260456 degrees of freedom), points show the result when areas 
with contributions Sgr/Low-Latitude stream/Virgo
are excised before carrying out the analysis (154336 degrees of freedom). 
}
\end{figure}

\begin{figure}[t]
\begin{center}
\hspace{-0.5cm}
\epsfxsize 9cm
\epsfbox{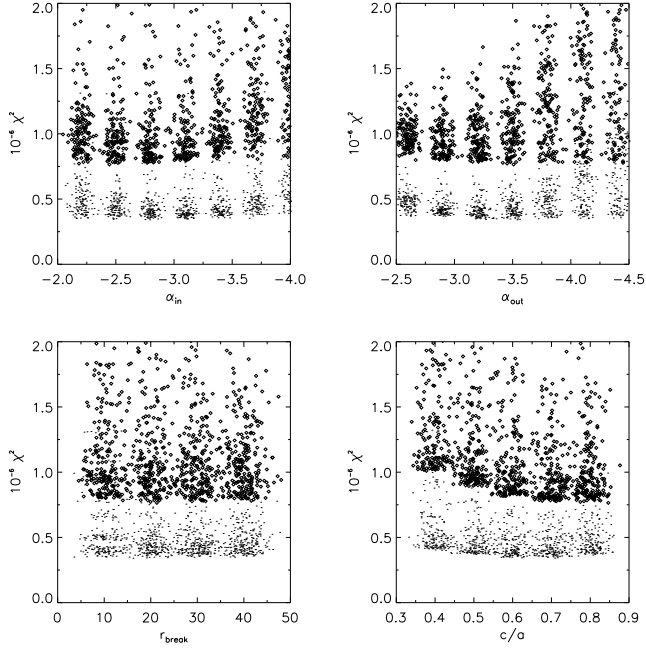}
\end{center}
\caption{\label{fig:parachi} 
The $\chi^2$ values of a large number of {\it oblate} halo
models.  The figure is formatted identically to Fig.\ \ref{fig:para}.
}
\end{figure}

\begin{figure}[t]
\begin{center}
\hspace{-0.5cm}
\epsfxsize 9cm
\epsfbox{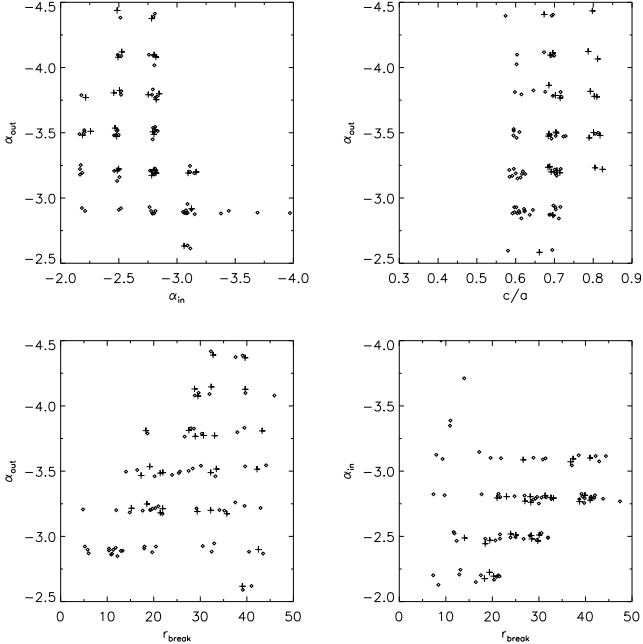}
\end{center}
\caption{\label{fig:para2} 
Covariance between different model parameters, for the `best' {\it oblate}
fits (diamonds show models with $\sigma$/total $<0.45$, whereas
crosses show models with $\chi^2 < 8\times 10^5$) 
for which all data with $b > 30\arcdeg$ were fit.  
Small random offsets are applied to the discrete values of 
$\alpha_{\rm in}$, $\alpha_{\rm out}$, $r_{\rm break}$ and $c/a$
to aid visibility.
}
\end{figure}

\begin{figure}[t]
\begin{center}
\hspace{-0.5cm}
\epsfxsize 9cm
\epsfbox{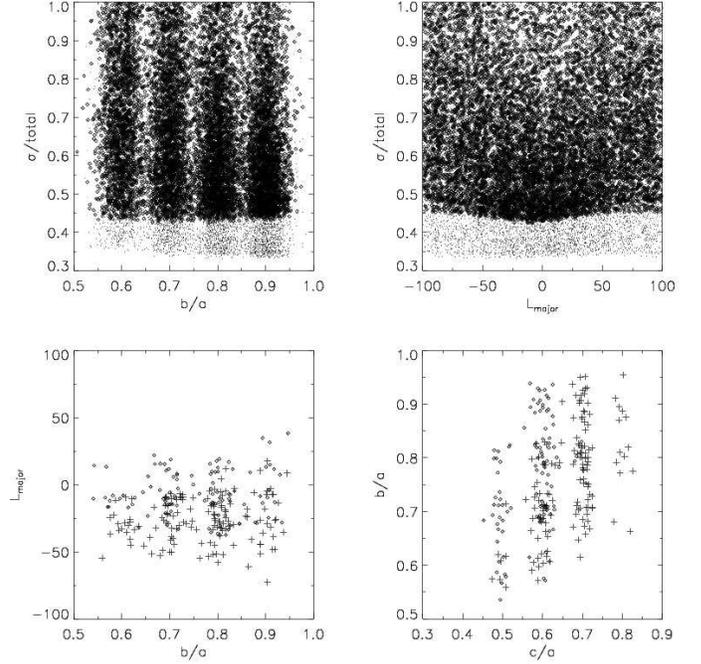}
\end{center}
\caption{\label{fig:tri} 
Data fitting results for the triaxial model halos, analogous to 
Fig.\ \ref{fig:para}.  In this figure, we show 
only the behavior of the `extra' parameters required
for a triaxial fit, as the behavior of 
$\alpha_{\rm in},\alpha_{\rm out},r_{\rm break}$ and $c/a$ is 
similar to the oblate case shown in Fig.\ \ref{fig:para}. 
Again, diamonds show the results for all data 
with $b>30\arcdeg$, and the points for the case where
Sagittarius, the Low-Latitude stream and the Virgo Overdensity
were masked out. 
The top two panels show how RMS depends on $b/a$ (where $b/a = 1$ is 
the oblate case and is not shown), and $L_{\rm major}$, the angle 
between the long axis of and the GC-Sun line.  
In the bottom two panels, we show covariance between $L_{\rm major}$ and 
$b/a$, and $b/a$ and $c/a$ for model fits with $\sigma$/total$<0.44$ (diamonds)
and $\chi^2 < 7.8 \times 10^5$ (crosses)
for which all data with $b > 30\arcdeg$ were been fit.  
Small random offsets are applied to the discrete values of 
$b/a$, $L_{\rm major}$ and $b/a$
to aid visibility.  Including triaxiality does not significantly
improve the quality of fit; when triaxiality is included
then values of $L_{\rm major}$ between $-$40 and 0 are favored, reflecting 
an attempt by the triaxial smooth halo model to fit out contributions from the 
Sagittarius tidal stream.
}
\end{figure}

\subsection{The `best fit' smooth oblate halo model} \label{sec:ob}

In Fig.\ \ref{fig:para}, we show how the residual
fraction depends on the halo parameters for a 
survey of parameter space for oblate halos.
It is immediately clear that these smooth models are a {\it very poor}
representation of the structure of the stellar halo, with 
values of $\sigma$/total $\ga 0.4$ for the best-fitting models 
for the case where all $b > 30\arcdeg$ data are fit, 
and $\sigma$/total $\ga 0.33$ for the case where 
Sagittarius, Virgo and the Low-Latitude overdensities are clipped.
Prolate models were attempted, and were all considerably poorer fits
than the oblate case shown here (i.e., the trend towards 
poorer fits in Fig.\ \ref{fig:para} with increasing
$c/a$ continues for $c/a > 1$).

In Fig.\ \ref{fig:hists}, we show with the black lines the distribution of the 
differences between observed and smooth model distributions 
in $0.5\arcdeg \times 0.5\arcdeg$ bins for both the case where
Sagittarius, Virgo and Low-Latitude structures were masked out (left
panel) and for all $b>30\arcdeg$ data (right panel).  In grey, we show
the distribution expected for Poisson noise around the smooth model 
alone.  The difference between the observed histograms and the Poisson 
expectation is the signal which we observe ($\sigma$/total $\sim 0.33, 0.43$ for the clipped and unclipped datasets, respectively)\footnote{Note that 
the appearance of Fig.\ \ref{fig:hists} depends on the 
adopted binning, through the contribution of Poisson uncertainties to the 
histogram of $\Delta \rho / \sigma$.  The value of $\sigma$/total 
is both in principle and in practice independent of 
binning scale.  Larger angular bins reduce the 
contribution of Poisson noise significantly, making the 
distribution of $\Delta \rho / \sigma$ significantly broader, 
while the value of $\sigma$/total is unchanged.  }.

From inspection of Fig.\ \ref{fig:para}, it is clear
that a variety of different combinations of parameters are able 
to provide similar values of $\sigma$/total.  The oblateness of the 
halo is best-constrained, with values of $c/a \sim 0.6$ 
preferred\footnote{The halo oblateness is affected
by the assumed value of $M_r$.  Variations of $M_r$ of $\pm0.5$\,mag
lead to changes in oblateness of $\mp0.1$.  Furthermore, if
the stellar halo has a binary fraction different from that of the
globular clusters used to calibrate the absolute magnitude and scatter
of turn-off stars, the values for absolute magnitude and 
scatter would be affected at the $\la 0.3$\,mag
level, leading to modest changes in recovered oblateness
\citep{larsen03}.}.  This determination of halo oblateness is in excellent 
agreement with that of previous work
\citep[e.g.,][]{chiba00,chen01,larsen03,lemon04,newberg,Ju06,xu06}.
Other parameters
are less well-constrained: various combinations
of $\alpha_{\rm in}$, $\alpha_{\rm out}$ and $r_{\rm break}$ are
capable of fitting the halo equally well.  Best fit stellar halo 
masses (over a radius range of 
1--40\,kpc) come out at $\sim 3.7\pm1.2 \times 10^8 {\rm M}_{\sun}$
for the models with $\sigma$/total $<0.45$, with 
considerable uncertainty from the mass-to-number ratio.

In Figs.\ \ref{fig:chisig} and 
\ref{fig:parachi} we show the relationship between 
$\chi^2$ and $\sigma$/total, and show the 
run of $\chi^2$ as a function of the smooth halo model 
parameters.  The minimum $\chi^2$ is $7.65 \times 10^5$ with 
260456 degrees of freedom for the case where all $b > 30\arcdeg$ data
are fit, and $3.41 \times 10^5$ with 154336 degrees of freedom for 
the case where Sagittarius, the low-latitude stream and Virgo are excised 
from the fit; in both cases, the probability of the data being drawn from 
the model are zero (to within floating point precision).  
One can see that $\sigma$/total and $\chi^2$ 
minimizations yield similar, but not identical results. 
The principal difference between $\sigma$/total
and $\chi^2$ minimization is that models with somewhat
higher $c/a \sim 0.7$ are preferred.  
This is because of the $1/\sigma^2$ weighting of $\chi^2$, 
that gives higher weight to better-populated pixels (in our
case, the pixels at larger radii; this tends to give Sagittarius
high weight in driving the fit).  Such a tendency towards higher
$c/a$ with at distances $\ga$20\,kpc has been claimed before
\citep{chiba00}; we do not comment further on this possible
trend here.   Nonetheless, the key message of these
plots is that minimization using $\chi^2$ and subsequent estimation 
of $\sigma$/total yields similar results, but with slightly larger values
of $\sigma$/total than our method, which chooses explicitly
to minimize the metric of interest in order to put a lower
limit on its value.  

The covariance of the different fitting parameters of 
the oblate case is illustrated 
in Fig.\ \ref{fig:para2}.  Models yielding $\sigma$/total $< 0.45$ are
shown as diamonds, and $\chi^2 < 8 \times 10^5$ as crosses, 
where all data with $b > 30\arcdeg$ are used.  It is clear
that the degeneracies in $\alpha_{\rm in}$, $\alpha_{\rm out}$ 
and $r_{\rm break}$ indicate that there are a number of different
ways to construct reasonable halo models 
(see \citealp{robin00} for similar results), with the general features
of a power law $\alpha_{\rm out} \sim -3$ in the 
outer parts and a similar or shallower power law
in the very inner parts of the halo at Galactocentric radii
$r_{\rm GC} \la 20$\,kpc\footnote{It is interesting in this context 
that there have been claims of 
a break in the power law of the stellar halo at $r_{\rm GC} \sim 20$\,kpc 
from counts of RR Lyrae 
stars (see \citealp{preston91}, although other analyses see no 
evidence for a break, e.g., \citealp{chiba00}).  }.  
It is important to note that the 
constraints on the `best fit' halo model are very weak, owing
to the significant degree of halo substructure.

\subsection{Triaxial models} \label{sec:tri}

The results for triaxial models are shown in 
Fig.\ \ref{fig:tri}.  We do not show the results
for the power-law parameters 
$\alpha_{\rm in}$, $\alpha_{\rm out}$ and $r_{\rm break}$, nor
the run of $\sigma$/total vs.\ $c/a$, as the results
for these parameters is very similar to the oblate halo case.
We focus instead on the results for the `new' parameters
$b/a$ and $L_{\rm major}$ (the angle between the long axis
and the line between the Galactic Center [GC] and Sun).  

The best triaxial fit is only very marginally better
than the best oblate fit, with $\sigma$/total$=0.42$; in particular,
the best triaxial fit is still a very poor fit to the 
stellar halo of the Milky Way as judged by either
$\sigma$/total or $\chi^2$.  Inspection of Fig.\ \ref{fig:tri}
shows that the best models are only mildly triaxial with
$b/a \ga 0.8$, and with $L_{\rm major} \sim -20$ (roughly lining up with the
Sagittarius stream).  In the bottom panels we show the covariance of the 
parameters of 
all models with $\sigma$/total$<0.44$ (diamonds) or $\chi^2 < 7.8 \times 10^5$
(crosses).  There is little obvious 
covariance between the `triaxiality' parameters, or between the 
power law parameters and the triaxiality parameters.  This 
stresses the difficulty in fitting a unique model to the halo;  
owing to the significant degree
of halo substructure, there are many ways to fit the halo by 
balancing problems in one part of the halo against a better
fit elsewhere. 

\subsection{A highly structured stellar halo}

The key point of this paper is that a smooth and symmetric (either
oblate/prolate or triaxial)
model is a poor representation 
of the structure of the stellar halo of our Milky Way.  
The $\sigma$/total of the $b > 30\arcdeg$ data around the model
is $>42$\%; even if the largest substructures are clipped, the
values of $\sigma$/total are $>33$\% (i.e., the largest
substructures contain $\sim$40\% of the total variance). 

One can obtain a visual impression of how poorly
fit the stellar halo is by a smooth model by 
examining Fig.\ \ref{fig:resid}, which shows the mean stellar density residuals
from the {\it best fit} oblate model.  The residuals
are smoothed by a $42'$ Gaussian kernel to suppress 
Poisson noise.  One can see that the residuals are highly structured
on a variety of spatial scales.  Particularly prominent 
are contributions from the well-known Sagittarius 
tidal stream (dominating all residuals for $20.5 \le r < 22.5$), 
the Low-Latitude stream (Galactic anticenter direction and $b < 35\arcdeg$), 
and the Virgo overdensity (particularly prominent in the $19.5 \le r < 20$
slice as the diffuse overdensity centered at $(l,b) \sim (280,70)$:
see \citealp{Ju06} and \citealp{newberg_new}).

\clearpage

There are a number of other less obvious structures.  
In the last three magnitude bins, one can discern the
`Orphan Stream' \citep{belokurov06_orp,grillmair06_orphan}, starting at $(l,b) \sim (250,50)$
and stretching to $(l,b) \sim (170,40)$ before 
disappearing into the noise (there is a clear distance gradient, such that 
as $l$ decreases the distance increases).  Visible also 
is a recently-identified structure of stars stretching from 
$(l,b) \sim (180,75)$ towards
$(l,b) \sim (45,45)$.  This structure, called 
the Hercules-Aquila Overdensity by \citet{belokurov07_her}, extends south 
of the Galactic plane (as shown in that paper) and 
is at a distance of $\sim 16$\,kpc from 
the Sun.  The Hercules-Aquila Overdensity is reflected as 
a distinct overdensity in 
color--magnitude space, shown in the lower
middle panel of Fig.\ \ref{fig:cmd}.  This CMD, obtained 
by subtracting a background field at $(l,b) \sim (15,45)$
from an overdensity field at $(l,b) \sim (44,40)$, 
shows a somewhat broadened MSTO with turn-off color
$g-r \sim 0.3$ (i.e., a similar color to the rest of the stellar halo).
Fig.\ \ref{fig:resid} illustrates that 
this very diffuse overdensity lies in a `busy' area of the halo, 
making its extent difficult to reliably estimate.
There are other potential structures visible, 
in particular in the most distant $22 \le r < 22.5$ bin. 
Some of the structure has low-level 
striping following the great circles along which the SDSS
scans\footnote{This striping has a modest effect on 
our measurement of $\sigma$/total, as illustrated in Fig.\ \ref{fig:dist}.
There are two main effects, working in counteracting directions: on one
hand, the striping will introduce a small amount of excess variance, 
on the other hand, galaxies misclassified as stars are smoothly
distributed across the sky, reducing the variance.  We chose to 
include the $22 \le r < 22.5$ bin in the analysis, noting that
its exclusion does not affect our results or conclusions.}, 
indicating that the structure is an artifact of uneven 
data quality in different stripes.  Other structures 
have geometry more suggestive of genuine substructure; 
we choose to not speculate on the reality (or `distinctness') 
of these structures at this stage
owing to the decreasing data quality at these faint limits.

\subsection{Structure as a function of distance}

\begin{figure}[t]
\begin{center}
\hspace{-1.cm}
\epsfxsize 8.5cm
\epsfbox{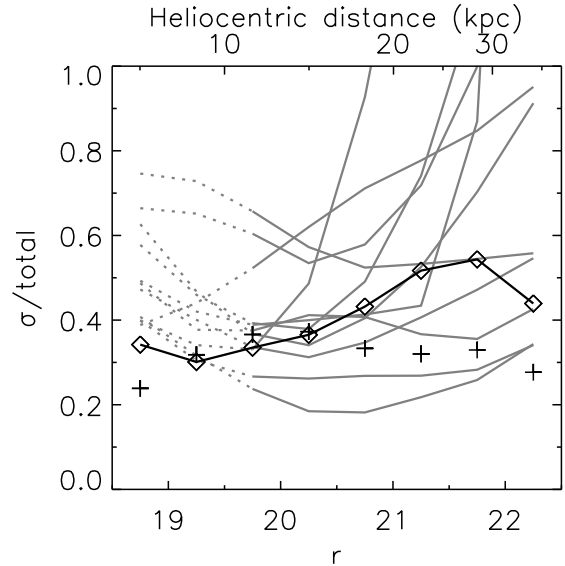}
\end{center}
\caption{\label{fig:dist} 
The substructure in the Milky Way stellar halo, compared to 
predictions from cosmological models.
The $\sigma$/total as a function of apparent magnitude (distance
assuming $M_r \sim 4.5$) for
the `best fit' oblate model.  Diamonds denote
the SDSS results for all $b > 30\arcdeg$ data; crosses
denote analogous results when the bulk of the Sagittarius 
and Low-Latitude tidal streams, and the Virgo
overdensity, have been excised from consideration.
The ensemble of solid gray lines show the predictions for  
$\sigma$/total from 11
models of stellar halo growth in a cosmological context
from \citet{bullock};  dotted 
lines are used at small radii where the simulations
are likely to be less robust.  In these simulations the entire halo 
arises, by model construction, from the disruption of satellite galaxies.
}
\end{figure}

The visual impression given by Fig.\ \ref{fig:resid} 
suggests an increasing
amount of deviation from a smooth halo at 
larger heliocentric distances.  We quantify
this in Fig.\ \ref{fig:dist}, where we show
the $\sigma$/total as a function of apparent magnitude
for all stars with $b > 30\arcdeg$ (diamonds). 
While it is clear that the exact values of $\sigma$/total
will depend somewhat on which smooth model happens to 
fit best, the value of $\sigma$/total doubles from distances
of $\sim 5$\,kpc to $\sim 30$\,kpc.  From comparison 
with the case when Sagittarius, the Low-Latitude
stream and the Virgo overdensity are removed before
calculation of the RMS, one can see that much 
of this increase in RMS is driven by the few large
structures; i.e., much of the RMS is contained in 
a few very well-defined structures at large radii.

\section{Comparison with expectations from a $\Lambda$CDM Universe}
\label{sec:disc}

\begin{figure*}[t]
\begin{center}
\hspace{-1.cm}
\epsfxsize 19.0cm
\epsfbox{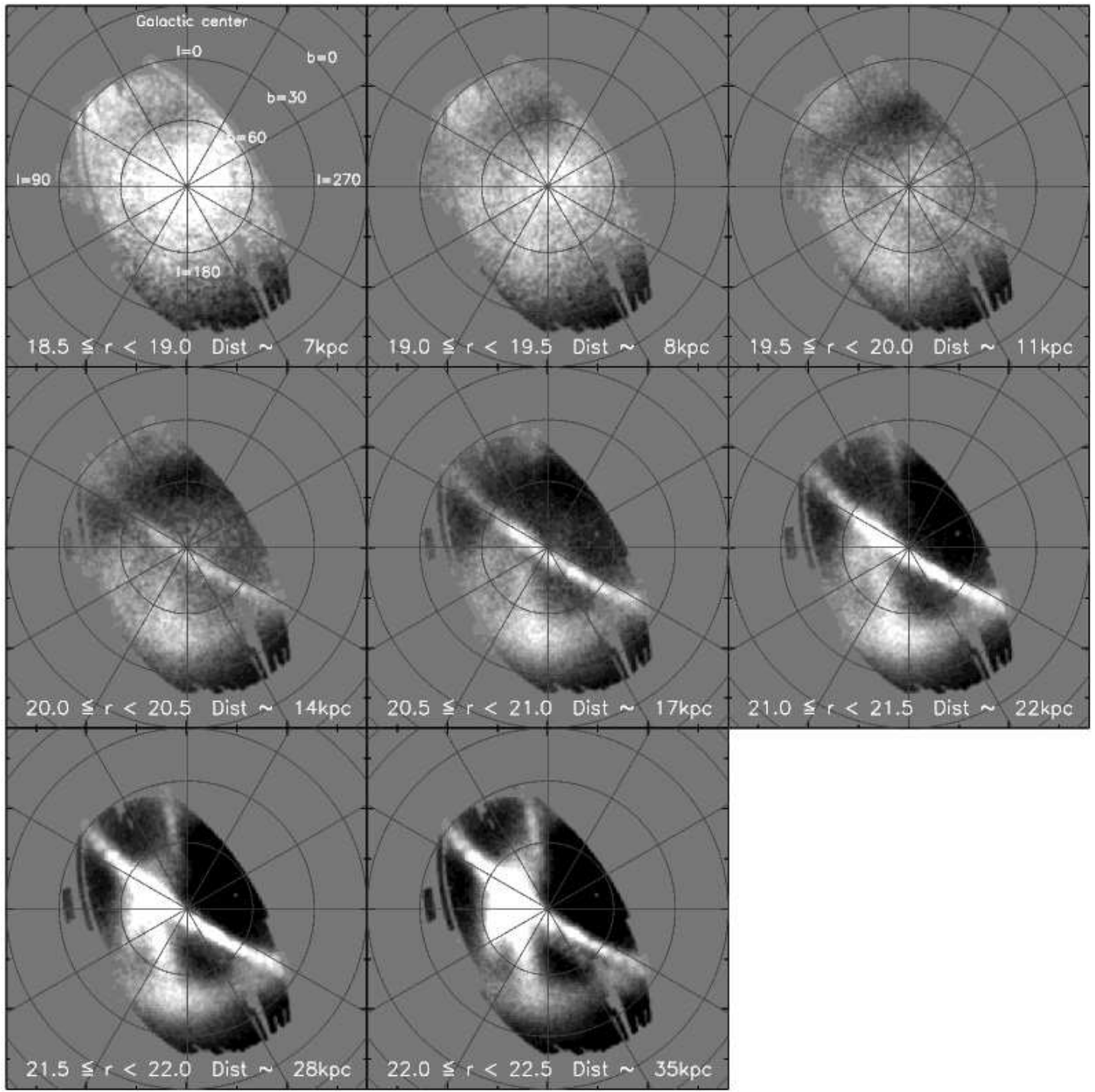}
\end{center}
\caption{\label{fig:model5} 
Residuals (simulation$-$smooth model) smoothed using a $\sigma = 42'$ Gaussian
from the best {\it oblate} model for Model 2 from \citet{bullock} 
in 8 different distance 
slices.  The gray scale saturates at $\pm 60$\% from the smooth model density.
}
\end{figure*}

\begin{figure*}[t]
\begin{center}
\epsfxsize 17.0cm
\epsfbox{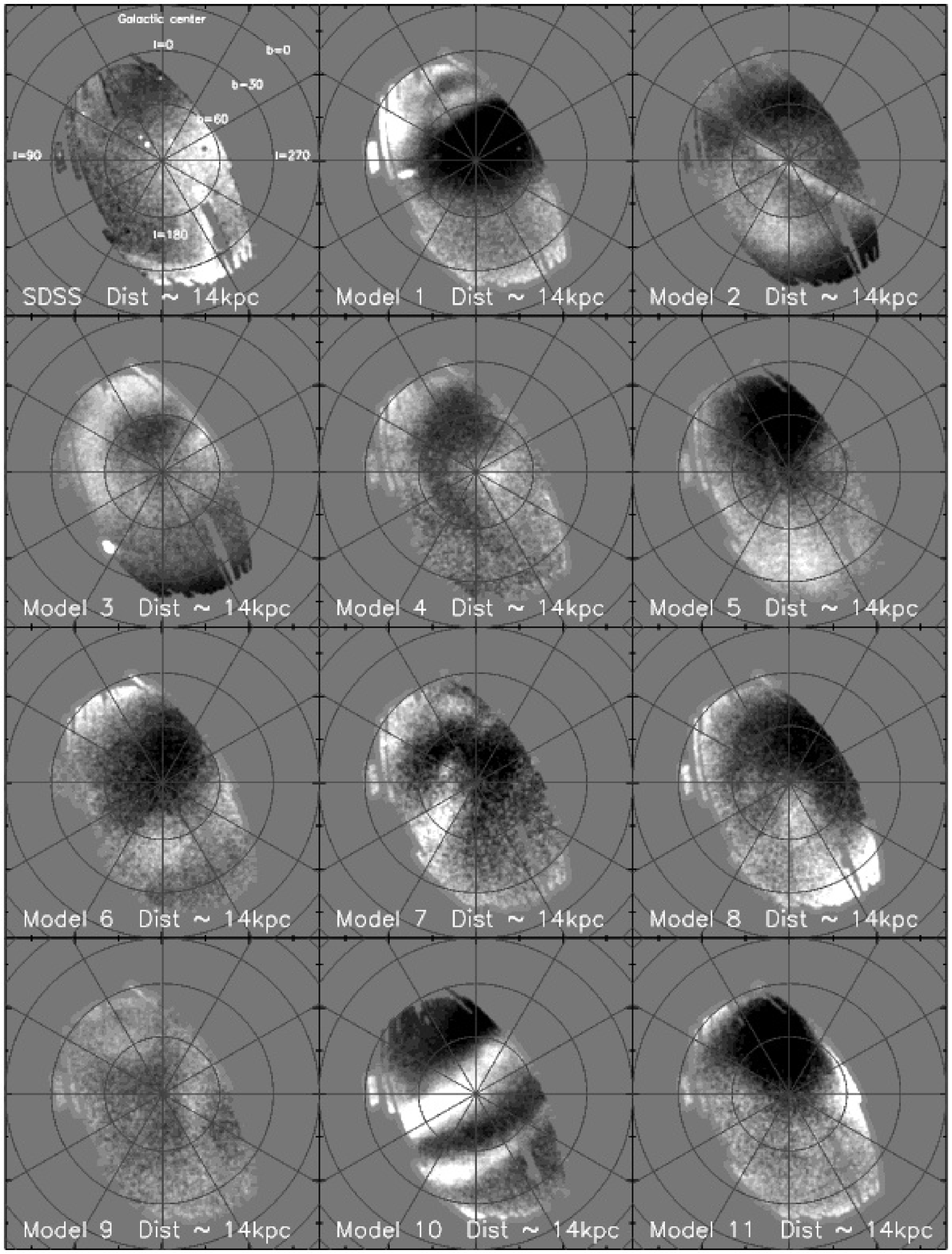}
\end{center}
\caption{\label{fig:models} 
Residuals (SDSS or simulations minus the smooth model)
smoothed using a $\sigma = 42'$ Gaussian from the best {\it oblate}
model fits for the SDSS data (top left panel) and for the 11 simulations
from \citet{bullock}.  We show only the $20 \le r < 20.5$ slice, 
corresponding to heliocentric distances $\sim 14$\,kpc.
The gray scale saturates at $\pm 60$\% from the smooth model density.
}
\end{figure*}

In this paper, we have attempted to fit smooth models to the stellar 
halo of the Milky Way.  Models containing 
$3.7 \pm 1.2 \times 10^8 {\rm M}_{\sun}$ in the 
radial range 1--40\,kpc with power-law density distributions
$\rho \sim r^{-3}$ were favored, {\it although all smooth 
models were a very poor fit to the data}.
We have found that
the stellar halo of the Milky Way halo is richly
substructured, with $\sigma$/total $\ga 0.4$.  The fractional amount of
substructure appears to increase with radius; this increase
is driven primarily by a few large structures.

To put our results into a cosmological context, we compare
the observations to predictions for stellar halo structure from appropriate 
models.  \citet{bkw} and \citet{bullock} studied the structure of 
stellar halos created {\it exclusively} through the merging and disruption
of reasonably realistic satellite galaxies.\footnote{\citet{abadi06} analyzed
the properties of the stellar halo of a disk galaxy formed
in a self-consistent cosmological simulation.  Such a self-consistent
simulation does not require that stellar halos be built up solely through
accretion; yet, the final halos produced were very similar to those
of \citet{bkw} and \citet{bullock}.}
These studies found that the debris
from disrupted satellite galaxies produced stellar halos with:
{\it i)} roughly power-law profiles with $\alpha \sim -3$
over 10--30\,kpc from the galactic center (e.g., Fig.\ 9 of 
\citealp{bullock}, see also \citealp{diemand05}, \citealp{moore06}), 
{\it ii)}  total stellar halo masses from 
$ \sim 10^9{\rm M}_{\sun}$ (integrated over all radii), and {\it iii)} 
richly substructured halos with increasingly
evident substructure at larger distances (e.g., Figs.\ 13 and 14 of 
\citealp{bullock}).  

\clearpage

\subsection{A quantitative comparison with simulated stellar halos}

\begin{figure}[t]
\begin{center}
\epsfxsize 9.0cm
\epsfbox{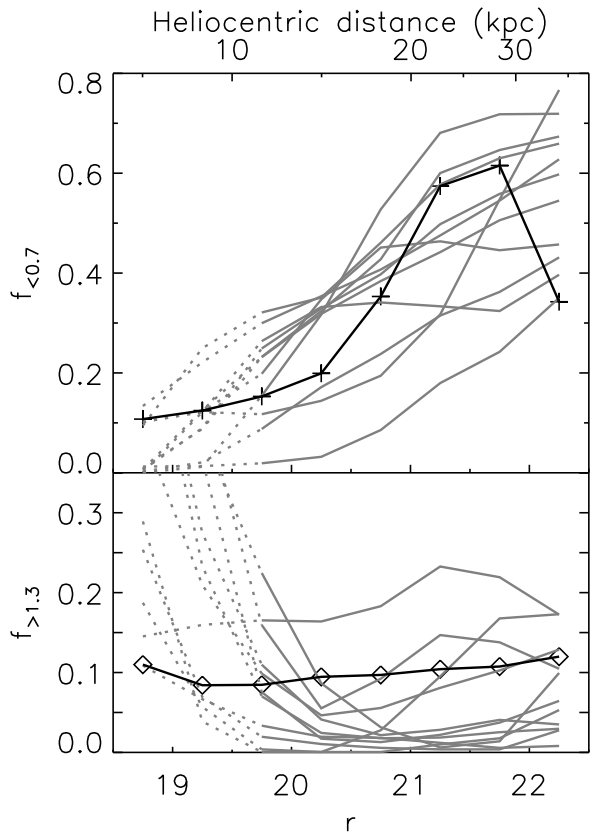}
\end{center}
\caption{\label{fig:underover} 
The fraction of the SDSS footprint in areas with $<70\%$ of the 
smooth model density (upper panel) and $>130\%$ of the smooth 
model density (lower panel) in both the observations (black line
with symbols) and the 11 simulations from \citet[gray lines]{bullock}.
Dotted lines denote where the simulations are argued to be less robust.
}
\end{figure}

We quantify the last statement through comparison of the 
SDSS data for the stellar halo with 11 simulated stellar halos 
from \citet{bullock}\footnote{The number of 
particles in the stellar halo of the \citet{abadi06} 
model galaxy was unfortunately too small to permit 
a proper comparison with the SDSS data.}.
These 11 simulated halos were generated at
random using semi-analytic
merger trees appropriate for a $\Lambda$CDM cosmology 
for a Milky-Way mass dark matter halo.
Maps of MSTO stars (analogous to our SDSS data) were constructed 
from the simulated $N$-body stellar halos, accounting for
all important observational effects, as follows.  The number of MSTO 
stars per particle was estimated using a 
ratio of 1 main sequence turn off star for every
8 $L_{\sun}$, as calibrated empirically using
Palomar 5.  MSTO stars were distributed in space by smoothing
over the 64 nearest $N$-body particle neighbors, using a Epanechnikov 
kernel of the form $(1-r^2)$.  Each star was assigned a 
simulated Galactic latitude, longitude, and heliocentric distance
assuming that the Sun is 8\,kpc from the Galactic center.  
The heliocentric distance is used to generate $r$-band 
apparent magnitudes for the MSTO stars
assuming an absolute magnitude $M_r = 4.5$ and 
scatter $\sigma_{M_r} = 0.9$ (following \S \ref{sec:models}).
The models were then placed 
in a Lambert equal-area projection, and 
the survey limits of the SDSS DR5 data analyzed in this paper
applied to the simulated maps.  
These simulations were analyzed
in the same way as the SDSS data, by fitting the same grid of oblate 
models.  The results are shown in Fig.\ \ref{fig:dist}, and 
Figs.\ \ref{fig:model5} and \ref{fig:models}.

Fig.\ \ref{fig:dist} shows the main result of this analysis:
all simulations predict a great deal of halo substructure, with 
values of $\sigma$/total $\ga 0.2$.  The typical smooth halo fitting 
parameters (where we quote the average and scatter derived from the 
fits to the 11 simulated stellar halos) 
are similar to that of the Milky Way's halo with 
$\alpha_{\rm out} \sim -3.4 \pm 0.6$, 
$M_{1<r/{\rm kpc}<40} \sim 2.8 \pm 1.5 \times 10^8 M_{\sun}$, 
and $c/a \sim 0.65 \pm 0.25$; values of $\alpha_{\rm in}$ within
$r_{\rm break} \sim 25$\,kpc 
tend to be higher than that observed for the Milky Way at $-1.3 \pm 0.7$. 
At small Galactocentric 
radii $\la 15$\,kpc, the simulations are expected to be much too 
structured (owing to the lack of a live Galactic potential, see 
\S 4.2 of \citealp{bullock}); accordingly, we show 
results for heliocentric distances $\la 10$\,kpc as dotted lines, and 
place little weight on the relatively high values of $\alpha_{\rm in}$ 
recovered by the best-fitting models.
At larger radii, where the simulation results are expected to be 
more robust, there are model halos with both less structure and 
more structure than the Milky Way's stellar halo.  
We illustrate this result in Figs.\ \ref{fig:model5} and 
\ref{fig:models}.  Fig.\ \ref{fig:model5} shows the 
residuals (simulation$-$best fit smooth
halo) for a model with very similar $\sigma$/total to the 
Milky Way on the same grey scale used for Fig.\ \ref{fig:resid}
in eight different apparent magnitude slices.  Fig.\ \ref{fig:models}
illustrates the diversity of simulated halos, showing 
the $20 \le r < 20.5$ apparent magnitude slice (corresponding to heliocentric
distances $\sim 14$\,kpc) for the SDSS and the 11 $\Lambda$CDM 
realizations of Milky Way mass stellar halos.  
A number of the general characteristics of the
simulations match the characteristics of the SDSS data:
the angular extent of `features' in the 
nearest bins is typically very large, whereas 
the angular width of streams in the distant bins tends
to be smaller.  In the distant bins, the halo substructure is
a combination of well-confined, relatively young streams and diffuse
sheets of stars from both older disruption events and young events
on almost radial orbits (K.\ Johnston et al., in preparation), 
with large-scale
overdensities and underdensities being seen.  

In Fig.\ \ref{fig:underover}, we explore the fraction of area in 
under- and overdensities in both the observations (black lines and symbols)
and the 11 $\Lambda$CDM realizations of 
Milky Way Mass stellar halos (gray lines).  
We quantify this by comparing the fraction of area for each apparent magnitude
slice with densities 30\% or more below the smooth model at that
radius ($f_{<0.7}$, shown in the upper panel), and the fraction 
of the area in each slice with densities 30\% or more above
the smooth model at that radius ($f_{>1.3}$, in the lower panel).  
This comparison is sub-optimal in the sense that both the model
and data have a non-zero contribution from Poisson noise (the immunity to 
Poisson noise was one
of the key advantages of the $\sigma$/total estimator), although 
we have reduced the Poisson noise by rebinning
the data and models in $4 \times 4$ pixel bins; with this rebinning, the
variance from counting uncertainties 
is 16 times smaller than the intrinsic variance.  
One can see the expected result that much of the sky 
area is covered in underdensities, with a smaller fraction of the 
sky in overdensities.  Again, the models at Galactocentric radii 
$\ga 15$\,kpc (where they are reliable) reproduce the general 
behavior of the observed stellar halo rather well.  Interestingly, 
the fraction of sky area in overdensities tends to be somewhat lower
in the models than in the observations (i.e., there may be room 
for the models to predict {\it more} substructure). 

This comparison shows that the the overall level of the substructure
seen in the Milky Way's stellar halo falls into the middle of the 
range of predictions from simulations --- simulations in which the 
stellar halo arises exclusively from the merging and disruption 
of satellite galaxies.  Furthermore, the character
of the structures in the simulated stellar halos
is very similar to those observed in the Milky Way\footnote{Model 2 in particular matches the trend in RMS with Heliocentric distance and the fraction 
in over/underdensities to within $\la 0.1$ for all bins with 
distance $\ga 10$\,kpc.}.
The models clearly have some shortcomings; in particular, the use of a
slowly-growing rigid potential 
for the central disk galaxy in the \citet{bullock} 
simulations leads to excess structure in the central parts. 
Furthermore, 
it is possible that the real stellar halo has a `smooth' component
formed either {\it in situ} in the potential well of the galaxy or accreted
so early that no spatial structure remains.
Our analysis shows that there is no need for such a smooth 
component to explain the data, and suggests that a smooth component
does not dominate the 
halo at radii $5 < r_{\rm GC}/$\rm kpc$ < 45$.
Yet, we have not tested quantitatively how large a smooth component could 
lie in this radial range and still lead to the observed RMS: such 
an exercise will be the object of a future work.

\subsection{Limitations of this comparison}

While there are steps which can and will be taken with this
dataset to sharpen the comparison with the simulations (e.g., 
a quantitative comparison of the morphology and spatial scale of 
substructure, and the investigation of substructure metallicities), 
it is nonetheless clear that `small number statistics' is 
a key limitation of this work.  The SDSS DR5 
contiguously covers only 1/5 of the sky, encompassing some 
5--10\% of all halo stars, with 
Galactocentric radii between 5 and 45\,kpc (as estimated by comparison 
of the smooth halo stellar masses with the actual mass 
contained in the maps).
Larger and deeper multi-color imaging
surveys will be required to expand the coverage of 
the Milky Way's stellar halo, probing to larger 
halo radii where models predict that halo substructure should be easier
to discern (see, e.g., the prominent substructures discovered by 
\citealp{sesar07} using 
RR Lyrae stars in the multiply-imaged `Stripe 82' of the SDSS).  
Yet, there is significant halo-to-halo scatter
in the simulated stellar halos; thus, matching 
the properties of a {\it single} stellar halo will always be a 
relatively easy task.  More powerful constraints will come
from studies of the stellar halos of statistical samples of galaxies
using high-resolution ground-based or HST data 
(see encouraging progress from e.g., \citealp{ferguson02} and 
\citealp{dejong07}).

\section{Conclusions}
\label{sec:conc}

In this paper, we have quantified the degree of (sub-)structure in 
the Milky Way's 
stellar halo.   We have used a sample of stellar
halo main sequence turn-off stars, isolated using 
a color cut of $0.2 < g-r < 0.4$, and fit oblate and triaxial 
broken power-law models of the density distribution 
to the data.

We find that the `best' fit oblateness of the stellar
halo is $0.5<c/a<0.8$ over the Galactocentric radial 
range 5 to 40\,kpc.
Other halo parameters are significantly 
less well-constrained; many different combinations of parameters
(including mild triaxiality)
can provide comparably good fits.  A single power law 
$\rho \propto r^{\alpha}$ with $\alpha = - 3$ provides
an acceptable fit.  Values of $-2 > \alpha > -4$ are 
also reasonable fits, as are halo profiles with somewhat shallower
slopes at $r \la 20$\,kpc and steeper slopes outside that 
range.  The halo stellar mass
of such models between Galactocentric radii of 
1 and 40\,kpc is $3.7 \pm 1.2 \times 10^8{\rm M}_{\sun}$, 
with considerable uncertainty from the conversion 
of the number of $0.2<g-r<0.4$ turn-off stars to mass.

Importantly, we find that {\it all} smooth models are very
poor fits to the spatial distribution of stellar halo stars.  
Deviations from smooth parameterized distributions, 
quantified using the RMS of the data around the 
model fit in $0.5\arcdeg \times 0.5\arcdeg$ bins ($>$100\,pc scales
at the distances of interest) give $\sigma$/total $\ga 0.4$, after
subtracting the (known) contribution of 
Poisson counting uncertainties.  Furthermore, the halo seems significantly
more structured at larger radii than in the inner $\sim 10$\,kpc;
a few individual structures dominate this increase in 
$\sigma$/total at larger radii.

Qualitatively, these results show that the stellar `substructure'
found in the Milky Way's halo is not at all a small perturbation 
on top of a smooth halo.  Remarkably, this same conclusion holds when
excising the most prominent known substructures from the 
analysis, such as the Sagittarius stream, and then considering the remaining
area of the sky.  

We compared these observational results with models of stellar 
halo growth in a cosmological context taken from \citet{bullock}.
In these models, the stellar halo arises exclusively from the disruption 
of and mergers with satellite galaxies.    
The models were analyzed in the same way as the SDSS data.  Their
models predict $\alpha \sim -3$ in the radial range 10--30\,kpc,
halo masses $\sim 10^9 {\rm M}_{\sun}$ integrated over all radii
(or masses $\sim 3 \times 10^8 {\rm M}_{\sun}$ in the radial 
range 1--40\,kpc),
and richly-structured stellar halos with $\sigma$/total $\ga 0.2$.
At radii where the model predictions are most robust, the models 
show a range of degrees of substructure, from substantially less
than that observed for the Milky Way to substantially more.  Furthermore, 
the character of the substructure appears very similar to 
that showed by the Milky Way's stellar halo.
While it is clear that the models are not perfect, this 
comparison lends considerable quantitative
weight to the idea that a dominant fraction of the stellar halo of the Milky
Way is composed of the accumulated debris from the disruption
of dwarf galaxies.  

\acknowledgements
We thank the anonymous referee for their excellent suggestions, 
and for encouraging exploration of the fraction of material in 
under- and overdensities.
We thank Jun-Hwan Choi and Martin Weinberg for useful discussions.
E.\ F.\ B.\ thanks the Deutsche Forschungsgemeinschaft for their support 
through the Emmy Noether Program.  D.\ B.\ Z.\ was supported
by a PPARC-funded rolling grant position.
V.\ B.\ was supported by a PPARC Fellowship. 
T.\ C.\ B. acknowledges partial support for this work from grant PHY 02-16783,
Physics Frontier Center/Joint Institute for 
Nuclear Astrophysics (JINA), awarded
by the US National Science Foundation.

Funding for the SDSS and SDSS-II has been provided by the Alfred P.
Sloan Foundation, the Participating Institutions, the National Science
Foundation, the U.S. Department of Energy, the National Aeronautics
and Space Administration, the Japanese Monbukagakusho, the Max Planck
Society, and the Higher Education Funding Council for England. The
SDSS Web Site is http://www.sdss.org/.
                                                                               
The SDSS is managed by the Astrophysical Research Consortium for the
Participating Institutions. The Participating Institutions are the
American Museum of Natural History, Astrophysical Institute Potsdam,
University of Basel, Cambridge University, Case Western Reserve
University, University of Chicago, Drexel University, Fermilab, the
Institute for Advanced Study, the Japan Participation Group, Johns
Hopkins University, the Joint Institute for Nuclear Astrophysics, the
Kavli Institute for Particle Astrophysics and Cosmology, the Korean
Scientist Group, the Chinese Academy of Sciences (LAMOST), Los Alamos
National Laboratory, the Max-Planck-Institute for Astronomy (MPIA), the
Max-Planck-Institute for Astrophysics (MPA), New Mexico State
University, Ohio State University, University of Pittsburgh,
University of Portsmouth, Princeton University, the United States
Naval Observatory, and the University of Washington.

\end{document}